\documentclass[twocolumn, 
  aps, pra,
  amsmath,amssymb,
  longbibliography,
  ]{revtex4-1}
\usepackage{graphicx,color}
\usepackage{amsmath}
\usepackage{natbib}
\usepackage{epsfig}
\begin{document}

\title{\color{blue} Vibrational model of thermal conduction for fluids with soft interactions}

\author{Sergey A. Khrapak}
\email{Sergey.Khrapak@gmx.de}
\affiliation{Joint Institute for High Temperatures, Russian Academy of Sciences, 125412 Moscow, Russia and \\Institut f\"ur Materialphysik im Weltraum, Deutsches Zentrum f\"ur Luft- und Raumfahrt (DLR), 82234 We{\ss}ling, Germany}

\begin{abstract}
A vibrational model of heat transfer in simple liquids with soft pairwise interatomic interactions is discussed. A general expression is derived, which involves an averaging over the liquid collective mode excitation spectrum. The model is applied to quantify heat transfer in a dense Lennard-Jones liquid and a strongly coupled one-component plasma. Remarkable agreement with the available numerical results is documented. A similar picture does not apply to the momentum transfer and shear viscosity of liquids.  
\end{abstract}

\date{\today}

\maketitle

\section{Introduction}

An accurate general theory of transport process in liquids is still lacking, despite considerable progress achieved over many decades~\cite{HansenBook,GrootBook,MarchBook}. As a result we often have to rely (when experimental data is not available) on phenomenological approaches, semi-quantitative models, and scaling relationships. In this context one can mention the activated jumps theory of self-diffusion in simple liquids~\cite{FrenkelBook}, the Stokes-Einstein relation between the self-diffusion and shear viscosity coefficients~\cite{FrenkelBook,ZwanzigJCP1983,Balucani1990,CostigliolaJCP2019,KhrapakMolPhys2019}, the excess entropy scaling of transport coefficients~\cite{RosenfeldPRA1977,DzugutovNature1996,RosenfeldJPCM1999,DyreJCP2018}, their freezing temperature scaling~\cite{RosenfeldPRE2000,OhtaPoP2000,VaulinaPRE2002,CostigliolaJCP2018,
KhrapakAIPAdv2018}, and many others. 

The focus of this paper is on thermal conduction in simple liquids. Perhaps the simplest expression for the thermal conductivity coefficient $\lambda$ is the Bridgman's expression~\cite{Bridgman1923} proposed about one century ago,
\begin{equation}\label{Bridgman}
\lambda=3c_{\rm s}n^{2/3},
\end{equation}         
where $c_{\rm s}$ is the sound velocity and $n$ is the density (we assume $k_{\rm B}=1$ and measure temperature in energy units throughout this paper). It can be derived by assuming that the atoms of the liquid are arranged in a cubic quasi-lattice with the mean interatomic separation $\Delta=n^{-1/3}$ and that the energy between quasi-layers perpendicular to the temperature gradient is transferred with the sound velocity $c_{\rm s}$. An additional assumption, that the heat capacity at constant volume of a monoatomic liquid 
is about the same as for a solid at high temperature and is given by the Dulong-Petit law, $c_V\simeq 3$, gives rise to the prefactor in Eq.~(\ref{Bridgman})~\cite{BirdBook}.        

Another simple model proposed by Horrocks and McLaughlin~\cite{Horrocks1960} considers the same idealization of the liquid structure, but assumes that the energy between successive layers is transferred due to vibrations with a characteristic Einstein frequency of the liquid's quasi-lattice, $\Omega_{\rm E}$. Omitting numerical coefficients of order unity, which contain geometrical factors and the probability that energy is transferred when two vibrating atoms collide, the coefficient of thermal conductivity is evaluated as
\begin{equation}\label{Horrocks}
\lambda = c_V \frac{\Omega_{\rm E}}{2\pi \Delta}.
\end{equation}

More recently Cahill and Pohl~\cite{Cahill1989,Cahill1992} re-analyzed the Einstein model of lattice heat conduction dating back to 1911 (and re-published in 2005~\cite{Einstein2005}). In Einstein's picture heat transport in crystals was a random walk of the thermal energy between neighboring atoms vibrating with random phases. Building on these ideas and assuming a Debye-like density of vibrational states, Cahill and Pohl proposed a so-called minimal thermal conductivity model~\cite{Cahill1989,Cahill1992}, which is in good
agreement with the measured thermal conductivities of many
amorphous inorganic solids, highly disordered crystals, and amorphous macromolecules~\cite{XiePRB2017}. In the high-temperature limit the model yields
\begin{equation}\label{Cahill}
\lambda\simeq 0.40n^{2/3}\left(c_l+2c_t\right),
\end{equation} 
where $c_l$ and $c_t$ are the longitudinal and transverse sound velocities, respectively. The applicability of this model to liquids was not discussed in Refs.~\cite{Cahill1989,Cahill1992}.

The purpose of this work is to put forward a generalization of the vibrational model of heat transfer in soft interacting particle liquids. A single general expression is derived, which reduces to Eq.~(\ref{Bridgman}), Eq.~(\ref{Horrocks}), or Eq.~(\ref{Cahill}) under special simplifying assumptions regarding the system vibrational properties. The reliability of the derived expression is then checked against recent numerical data on the thermal conductivity coefficient of a dense Lennard-Jones (LJ) liquid  and a strongly coupled one-component plasma (OCP) fluid. It is demonstrated that the model describes accurately the numerical data with no free parameters. Relations between the coefficients of thermal conductivity and viscosity in the liquid state are briefly discussed. It is shown that the mechanisms of momentum and heat transfer in liquids are different.

\section{Model}

\begin{figure}
\includegraphics[width=8.cm]{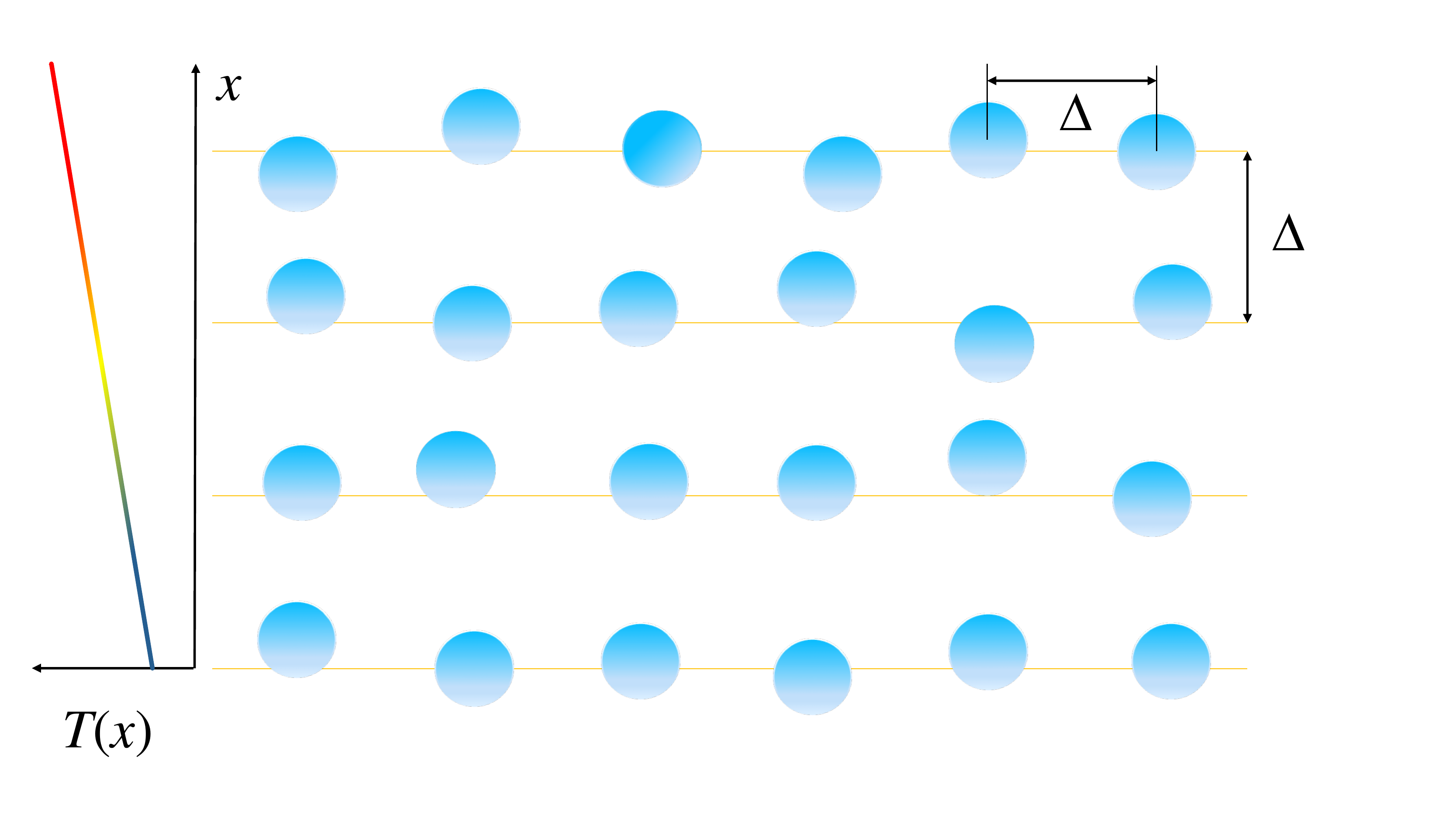}
\caption{(Color online) Two-dimensional slice of a quasi-layered three-dimensional fluid model under consideration. The average inter-particle separation within one quasi-layer and between quasi-layers is $\Delta = n^{-1/3}$. The temperature increases from bottom to top.}
\label{Fig1}
\end{figure}

Similar to the approaches by Bridgman and Horrocks and McLaughlin, a liquid is approximated by a layered structure with layers perpendicular to the temperature gradient and separated by the distance $\Delta = n^{-1/3}$. The particle density in each such quasi-layer is $\Delta^{-2}$. A sketch of the considered idealization is shown in Fig.~\ref{Fig1}. In contrast to the situation in a crystalline solid, the atomic positions in the liquid's quasi-layers are not fixed. The atoms can migrate within one layer as well as between different layers, however, the time scale of these migrations is relatively long. Considering the activated jump theory of self-diffusion in the liquid state, the following picture can be adopted. An atom oscillates almost harmonically about a local equilibrium position (determined by the interaction with other atoms), until it suddenly finds a ``free'' place amongst its nearest neighbours and jumps there, rearranging equilibrium positions of the neighbouring atoms. An important point is that the average waiting time between such rearrangements is much longer than the period of oscillations in a temporal equilibrium position. This picture is clearly more appropriate for sufficiently soft interatomic interactions and much less adequate for hard-sphere-like systems.      

Now, if a temperature gradient is applied, the average difference in energy between the atoms of adjacent layers is $\Delta (dU/dx)$, where $U$ is the internal energy. In the considered model, the energy between successive layers is transferred when two vibrating atoms from adjacent layers ``collide'' (this should not be a physical collision; the atoms just need to approach by a distance that is considerably shorter than the average interatomic separation). The characteristic vibrational frequency of the liquid's quasi-lattice is $\nu$ and this defines the characteristic energy relaxation frequency, according to Einstein's picture~\cite{Cahill1992}. Then, the energy flux per unit area is
\begin{equation}\label{flux}
\frac{dQ}{dt}=-\frac{\nu}{\Delta} \frac{dU}{dx},
\end{equation}  
where the minus sign indicates that the heat flow is down the temperature gradient. On the other hand, Fourier's law for the heat flow reads
\begin{equation}\label{Fourier}
\frac{dQ}{dt}=-\lambda\frac{dT}{dx},
\end{equation}
where $\lambda$ is the thermal conductivity coefficient, which is a scalar in isotropic liquids. Combining Eqs.~(\ref{flux}) and (\ref{Fourier}) we immediately get
\begin{equation}\label{Cond1}
\lambda=\frac{dU}{dT}\frac{\nu}{\Delta}=c_V\frac{\nu}{\Delta}=c_V\frac{\langle \omega \rangle}{2\pi \Delta}.
\end{equation}
It has been implicitly assumed that the characteristic frequency of energy exchange is equal to the average vibrational frequency of an atom, $\nu=\langle \omega \rangle/2\pi$ (which is a factor of two smaller than in the Cahill and Pohl model~\cite{Cahill1992}). The remaining step is to evaluate this average vibrational frequency. Since the actual frequency distribution can be quite complex in liquids, and can vary from one type of liquid to another, some simplifying assumptions have to be employed.

In the simplest Einstein approximation all atoms vibrate with the same (Einstein) frequency $\Omega_{\rm E}$ (on time scales shorter than the rearrangement waiting time) and, hence, $\langle \omega \rangle = \Omega_{\rm E}$. We recover immediately the expression by Horrocks and McLaughlin, Eq.~(\ref{Horrocks}).

As an improvement, let us consider a Debye spectrum, characterized by the vibrational density of states that is proportional to $\omega^2$, $g(\omega)\propto \omega^2$. We get
\begin{equation}
\langle \omega \rangle=\int_0^{\omega_{\rm D}}g(\omega)\omega d\omega \left[\int_0^{\omega_{\rm D}}g(\omega)d\omega\right]^{-1}=\frac{3}{4}\omega_{\rm D},
\end{equation}
where $\omega_{\rm D}$ is the cutoff Debye frequency. The latter can be estimated from the condition 
\begin{displaymath}
\Omega_{\rm E}^2=\langle \omega^2 \rangle = \frac{3}{5}\omega_{\rm D}^2,
\end{displaymath}
which yields $\langle \omega \rangle= 0.968 \Omega_{\rm E}$, which is again close to the result by Horrocks and McLaughlin.  

As an alternative, we can use an acoustic spectrum, $\omega=c_{\rm s}k$, supplemented by an appropriate cut-off of the wave numbers $k_{\rm max}$. Then, the standard averaging procedure results an analogue of the Bridgman equation (\ref{Bridgman}), to within a numerical coefficient. 

As a more general approximation, assume that a dense liquid support one longitudinal and two transverse modes. Averaging in $k$-space yields  
\begin{equation}\label{integral}
\langle \omega \rangle = \frac{1}{6\pi^2 n}\int_0^{\rm k_{\rm max}}k^2dk\left[\omega_l(k)+2\omega_t(k)\right],
\end{equation} 
where the cutoff $k_{\rm max}$ is chosen to provide $n$ oscillations for each collective mode, so that
\begin{displaymath}
\frac{4\pi}{3}\left(\frac{k_{\rm max}}{2\pi}\right)^3=n.
\end{displaymath}
If we deal with acoustic-like dispersion relations,
\begin{displaymath}
\omega_l(k)\simeq c_l k,\quad\quad \omega_t(k)\simeq c_tk,
\end{displaymath}  
the integration is trivial and we immediately get
\begin{equation}\label{Cahill1}
\lambda \simeq \frac{1}{4}\left(\frac{3}{4\pi}\right)^{1/3} c_V n^{2/3}\left(c_l+2c_t\right). 
\end{equation}
If we additionally assume $c_V\simeq 3$ the results similar to that of Cahill and Pohl is recovered, but with a slightly larger numerical coefficient, $\lambda \simeq 0.47 n^{2/3}\left(c_l+2c_t\right)$.

Thus, all three simple expressions appearing in the Introduction can be considered as just special cases of the more general expression (\ref{Cond1}). Note also that averaging in Eq.~(\ref{integral}) can be applied to systems with deviations from the acoustic dispersion. A remarkable example corresponding to the OCP fluid will be considered later in this paper.     

An important remark should be made before we conclude this Section. In equation (\ref{integral}) the integration over $k$ is performed all the way from zero to $k_{\rm max}$ for both the longitudinal and transverse modes. In this way the existence of a $k$-gap for the transverse collective mode is not taken into account. This ``$k$-gap'' implies a minimum (critical) wave number, $k_*$, below which transverse (shear) waves cannot propagate, which is a
well known property of the liquid state~\cite{HansenBook,GoreePRE2012,BolmatovPCL2015,TrachenkoRPP2015,KhrapakJCP2019,
KryuchkovSciRep2019}. However, since the contribution from the small $k$-region to the integral in Eq.~(\ref{integral}) is not essential ($\langle \omega\rangle \propto \int k^3dk$), the existence of the $k$-gap is not significant as long as $k_*\ll k_{\rm max}$. As the liquid temperature increases and (or) density decreases the $k$-gap widens and should be properly accounted for. However, in this regime the applicability of the vibrational model itself becomes questionable so we do not elaborate on this further. It should be additionally mentioned that since the $k$-gap width is directly related to the magnitude of $c_V$~\cite{KryuchkovPRL2020},  the $k$-gap and $\lambda$ are nevertheless implicitly related. 

In the following we verify the proposed model against the available results on heat conduction in a dense LJ liquid and in a strongly coupled OCP fluid.

\section{Lennard-Jones liquid}    

The Lennard-Jones potential, which is often used to approximate interactions in liquefied noble gases, reads 
\begin{equation}
\phi(r)=4\epsilon\left[\left(\frac{\sigma}{r}\right)^{12}-\left(\frac{\sigma}{r}\right)^{6}\right], 
\end{equation}
where  $\epsilon$ and $\sigma$ are the energy and length scales (or LJ units), respectively. The density, temperature, pressure, and energy expressed in LJ units are $n_*=n\sigma^3$, $T_*=T/\epsilon$, $p_*= P\sigma^3/\epsilon$, and $u_*=U/N\epsilon$. 

The LJ system is one of the most popular and extensively studied model systems in condensed mater physics. Many results on transport properties have been published over the years. Here we use the numerical results by Meier, who tabulated very accurate thermal conductivity coefficients along a close-critical isotherm $T_*=1.35$~\cite{Meier2002}. These results are particularly suitable in the present context, because in addition to the thermal conductivity coefficient, the data for the thermodynamic properties (i.e. specific heat $c_V$, the reduced energy $u_*$ and the reduced pressure $p_*$) as well as other transport coefficients (diffusion, shear and bulk viscosities) for investigated state points were also tabulated. This is a rare case when all the necessary information to perform a detailed comparison is immediately at hand.

Both the longitudinal and transverse modes of the LJ liquid exhibit an acoustic-like dispersion and therefore Eq.~(\ref{Cahill1}) is used for comparison. 
The sound velocities $c_l$ and $c_t$ were not tabulated in Ref.~\cite{Meier2002}, but they can be easily expressed (for the LJ system) in terms of the system pressure and energy~\cite{Meier2002,ZwanzigJCP1965,KhrapakMolecules2020}, and this is how they have been evaluated.             

\begin{figure}
\includegraphics[width=8.cm]{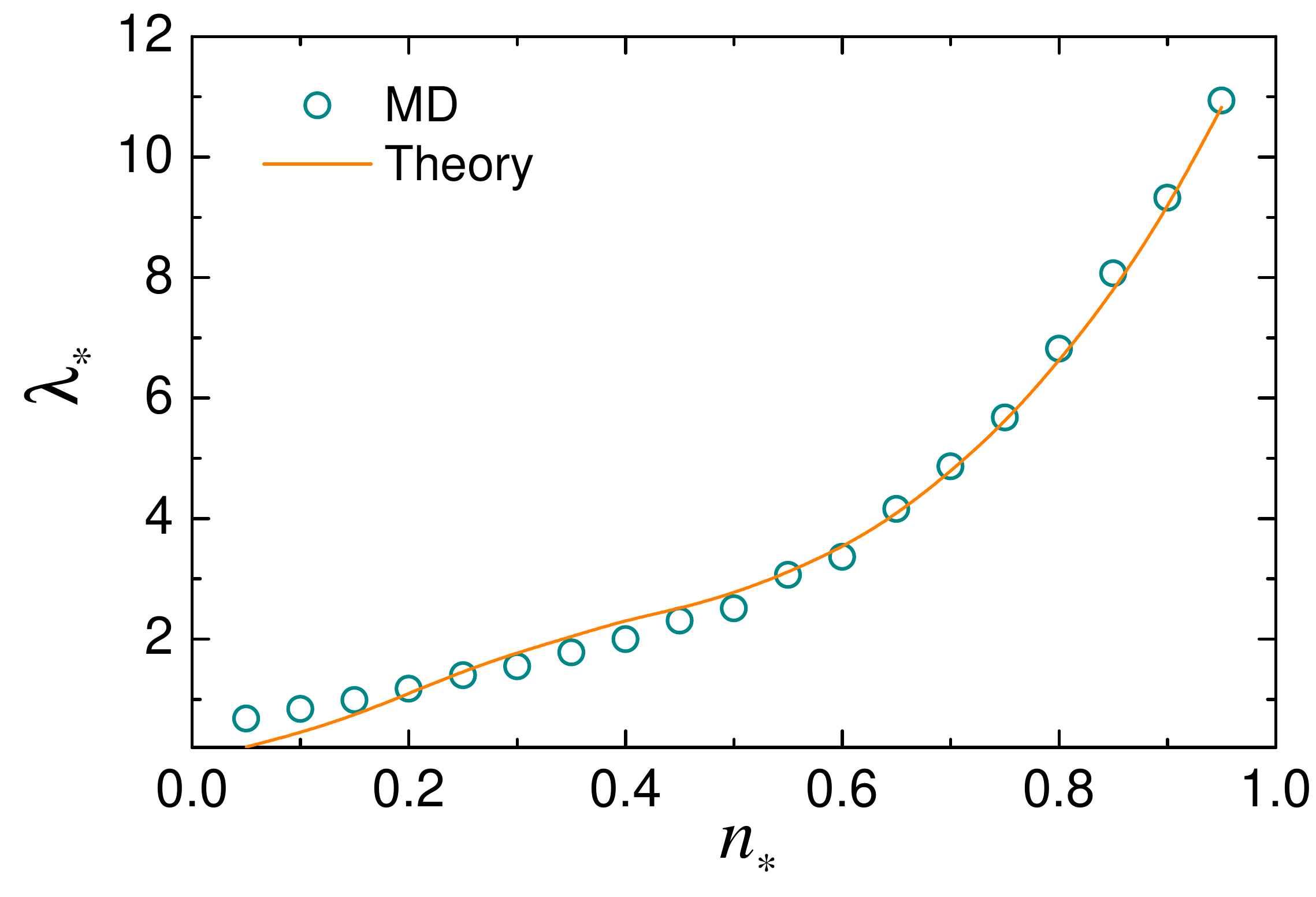}
\caption{Reduced thermal conductivity coefficient $\lambda_*$ versus the reduced density $n_*$ of a Lennard-Jones fluid along an isotherm $T_*=1.35$. Symbols correspond to numerical results from Ref.~\cite{Meier2002}. The solid curve is calculated using Eq.~(\ref{Cahill1}).}
\label{Fig2}
\end{figure}

The comparison beteen the vibrational model of Eq.~(\ref{Cahill1}) and numerical results from Ref.~\cite{Meier2002} is shown in Fig~\ref{Fig2}. The heat transport coefficient is made dimensionless using the LJ units: $\lambda_* = \lambda \sigma^2\sqrt{m/\epsilon}$. The agreement is excellent in the dense liquid regime, but becomes poor at low densities ($n_*\lesssim 0.2$) as can be expected. Note that in the vicinity of the critical density, some (modest) critical enhancement of the thermal conductivity coefficient is reproduced in both simulation and theory.  

\section{One-component plasma}

The OCP model is an idealized
system of point charges immersed in a neutralizing 
uniform background of opposite charge (e.g. ions in the immobile background of electrons or vice versa)~\cite{BrushJCP1966,deWitt1978,BausPR1980,IchimaruRMP1982}. This
model is of considerable practical interest as it is relevant to a wide class of physical systems, including for example laboratory and space plasmas, planetary interiors, white dwarfs, liquid metals, and electrolytes. There are also relations to various soft matter systems such as charged colloidal suspensions and complex (dusty) plasmas~\cite{FortovUFN,FortovPR,FortovBook}. From the fundamental point of view OCP is characterized by a very soft and long-ranged Coulomb interaction potential, $\phi(r)= e^2/r$, where $e$ is the electric charge. This potential is much softer than the Lennad-Jones potential considered above and this results in important differences regarding the collective mode properties. For this reason OCP represents a very important reference system to verify the validity of the vibrational model of thermal conductivity discussed here.      


Transport properties of the OCP and related system are very well investigated in classical MD simulations. Extensive data on the self-diffusion~\cite{DaligaultPRL2006,DaligaultPRL2012,DaligaultPRE2012,
KhrapakPoP2013}, shear viscosity~\cite{DonkoPoP2000,SalinPRL2002,BasteaPRE2005,DaligaultPRE2014,KhrapakAIPAdv2018}, and thermal conductivity~\cite{DonkoPoP2000,DonkoPRL1998,DonkoPRE2004,ScheinerPRE2019} have been published and discussed in the literature. Substantial progress in ab initio studies of related systems has been also achieved~\cite{FrenchNJP2019,FrenchAstroJ2019}. A large collection of data on the shear viscosity of strongly coupled plasmas has been analysed in connection to the lower bound on the ratio of the shear viscosity coefficient to the entropy density, obtained using string theory methods~\cite{FortovPRL2013}.      

Before we proceed further, let us quickly summarize some important properties of the OCP. The particle-particle correlations and thermodynamics of the OCP are characterized by a single dimensionless coupling parameter $\Gamma=e^2/aT$, where $a=(4\pi n/3)^{-1/3}$ is the Wigner-Seitz radius, and $T$ is the temperature in energy units ($k_{\rm B}=1$). The coupling parameter essentially plays the role of an inverse temperature or inverse interatomic separation. In the limit of
weak coupling (high temperature, low density), $\Gamma\ll 1$, the OCP is in a disordered gas-like state. Correlations increase with coupling and, at $\Gamma\gtrsim 1$, the OCP exhibits properties characteristic of a fluid-like phase (low temperature, high density). The fluid-solid phase transition occurs at $\Gamma\simeq 174$~\cite{IchimaruRMP1982,DubinRMP1999,KhrapakCPP2016}.

The dynamical properties of the OCP are determined by the plasma frequency $\omega_{\rm p}=\sqrt{4\pi e^2 n/m}$, which plays the role of the inverse time scale. All other important frequencies are proportional to $\omega_{\rm p}$. For example, the Einstein frequency $\Omega_{\rm E}$ can be quite generally expressed using the pairwise interaction potential $\phi(r)$ and the radial distribution function $g(r)$~\cite{BalucaniBook}:
\begin{equation}
\Omega_{\rm E}^2=\frac{n}{3m}\int d{\bf r}\nabla^2\phi(r)g(r).
\end{equation}
In the OCP case, $g(r)$ should be substituted by $g(r)-1$ due to the presence of the neutralizing background. The potential satisfies $\nabla^2\phi(r)=-4\pi e^2\delta({\bf r})$, by virtue of the Poisson equation. From this we immediately get the familiar identity $\Omega_{\rm E}=\omega_{\rm p}/\sqrt{3}\simeq 0.577\omega_{\rm p}$.

The actual spectrum of OCP collective excitations is different from that of a LJ liquid. At sufficiently strong coupling we also have one longitudinal and two transverse modes. However, the longitudinal mode does not exhibit the acoustic-like dispersion, it is a plasmon mode  (for an example of numerically computed and analytical dispersion relations see e.g. Refs.~\cite{GoldenPoP2000,SchmidtPRE1997,KhrapakIEEE2018}). We approximate the long-wavelength dispersion relations of these modes with
\begin{equation}\label{dispersion}
\omega_l^2\simeq \omega_{\rm p}^2-c_{l}^2k^2, \quad \omega_t^2\simeq c_{t}^2k^2.
\end{equation} 
Here $c_l$ is not the true acoustic velocity, this notation is only kept for simplicity. From the identity $\omega_{l}^2+2\omega_{t}^2=\omega_{\rm p}^2$ (valid for the OCP system) we get $c_l^2=2c_t^2$. The quantities $c_l$ and $c_t$ can be evaluated using the quasi-localized charge approximation, where they appear as functions of the reduced excess energy~\cite{GoldenPoP2000,KhrapakPoP2016} 
\begin{equation}\label{c}
c_l^2=-\frac{4}{15}v_{\rm T}^2u_{\rm ex}, \quad c_t^2=-\frac{2}{15}v_{\rm T}^2u_{\rm ex},
\end{equation}
where $u_{\rm ex}=U/NT-3/2$ (note that $u_{\rm ex}$ is negative at strong coupling due to the presence of the neutralizing background).  For $u_{\rm ex}$ we use a simple three-term equation proposed in Ref.~\cite{KhrapakPoP10_2014}, based on extensive Monte Carlo simulation data from Ref.~\cite{CaillolJCP1999},
\begin{equation}\label{EoS}
u_{\rm ex}\simeq -\frac{9}{10}\Gamma +0.5944\Gamma^{1/3}-0.2786.
\end{equation} 
The first term corresponds to the so-called ion sphere model (ISM)~\cite{IchimaruRMP1982,DubinRMP1999,KhrapakPoP2014}, which provides a dominant contribution at strong coupling. If we keep only this term in the expressions for $c_l$ and $c_t$ in Eq.~(\ref{c}), then the integration in Eq.~(\ref{integral}) results in $\langle \omega \rangle \simeq 0.523\omega_{\rm p}$. It was verified that keeping terms beyond ISM does not lead to any appreciable deviations from this result in the strongly coupled regime.

It should be pointed out that the kinetic terms (i.e. the Bohm-Gross term in the plasmon dispersion and a similar term in the transverse dispersion) are not included in Eq.~(\ref{dispersion}). These are numerically small at strong coupling and can be safely neglected. From a pragmatic point of view this is well justified away from the point where the negative dispersion of the plasmon mode sets in ($d\omega/dk<0$ at $k\rightarrow 0$). The onset of negative dispersion takes place at $\Gamma\simeq 10$~\cite{HansenJPL1981,MithenAIP2012,KorolovCPP2015,KhrapakOnset} and this limits the applicability of the approach from the side of weak coupling (in addition to neglecting $k$-gap in the transverse mode).   

\begin{figure}
\includegraphics[width=8.cm]{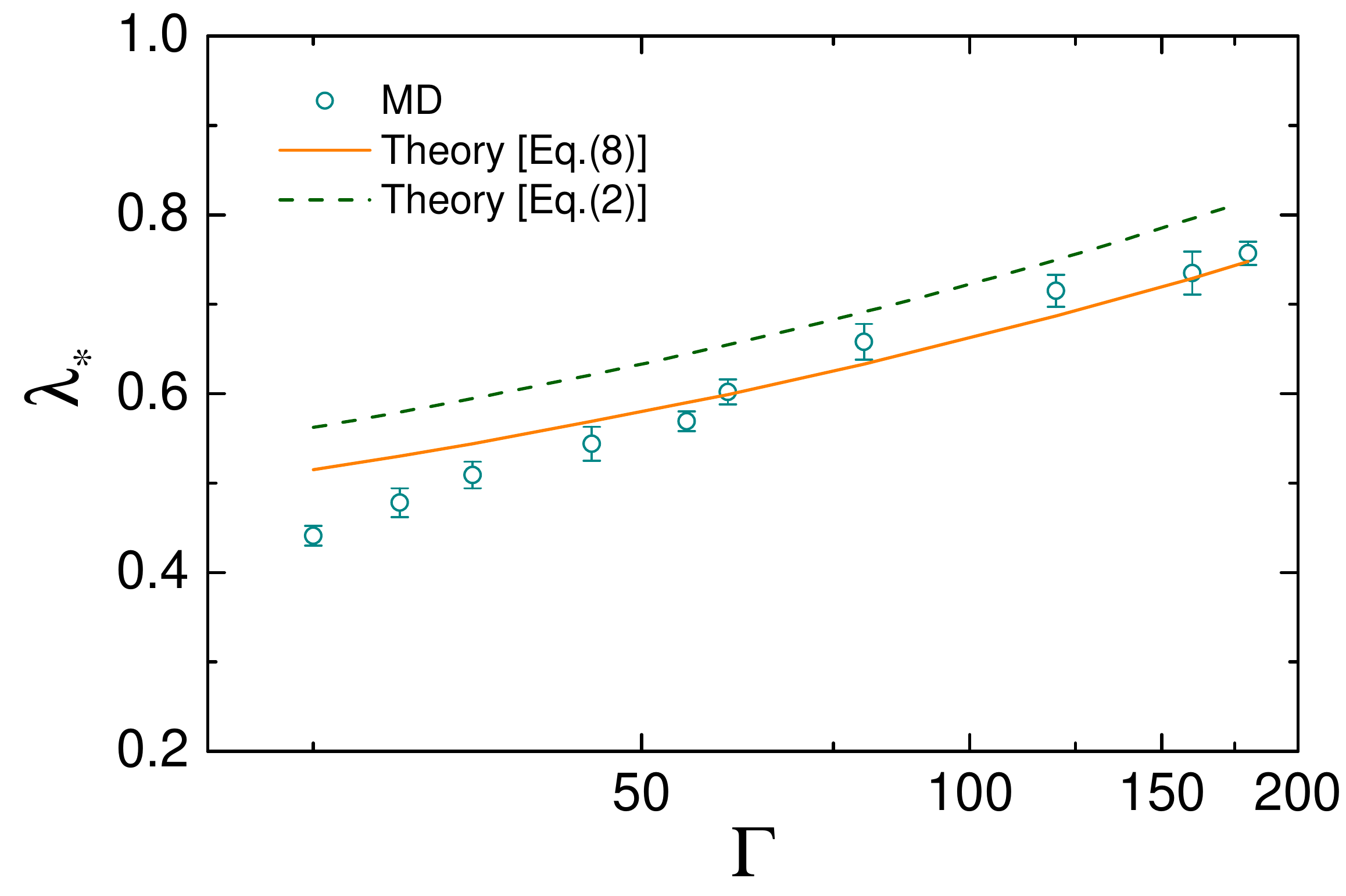}
\caption{Reduced thermal conductivity coefficient $\lambda_*$ of a strongly coupled OCP versus the coupling parameter $\Gamma$. Symbols correspond to numerical results from Ref.~\cite{ScheinerPRE2019}. Curves are calculated using the vibrational model discussed in this work: Solid curve -- Eqs.~(\ref{Cond1}) and (\ref{integral}); Dashed curve -- Eq.~(\ref{Horrocks}).}
\label{Fig3}
\end{figure}

The main dependence on $\Gamma$ in the strongly coupled regime is expected from the variation of $c_V$ with $\Gamma$. Expressing conventional thermodynamic identities in terms of $\Gamma$~\cite{KhrapakPRE03_2015,KhrapakPoP2015} we get  
\begin{equation}
c_V(\Gamma)=\frac{3}{2}+u_{\rm ex}-\Gamma\frac{\partial u_{\rm ex}}{\partial \Gamma}\simeq \frac{3}{2}+0.3936\Gamma^{1/3}-0.2786,
\end{equation}
where the EoS of Eq.~(\ref{EoS}) has been employed.

The theoretical model is compared with the numerical results from Ref.~\cite{ScheinerPRE2019} in Fig.~\ref{Fig3}. Following  the standard plasma physics nomenclature, the reduced thermal conductivity coefficient is defined as $\lambda_*=\lambda/n\omega_{\rm p}a^2$.
Two theoretical curves are plotted. The solid one corresponds to the averaging using Eqs.~(\ref{Cond1}),(\ref{integral}), and (\ref{dispersion}). The dashed curve is plotted using a simpler Eq.~(\ref{Horrocks}). The theoretical curves are relatively close, the former curve demonstrates better agreement as could be expected. Overall, the agreement between the theory and simulation in the strongly coupled regime, $\Gamma\gtrsim 50$ is remarkably good, especially taking into account the absence of free parameters. For weaker coupling the model overestimates the thermal conductivity coefficients, but its applicability becomes questionable there for the reasons discussed above.
     
\section{Shear viscosity} 

\begin{figure}
\includegraphics[width=8.cm]{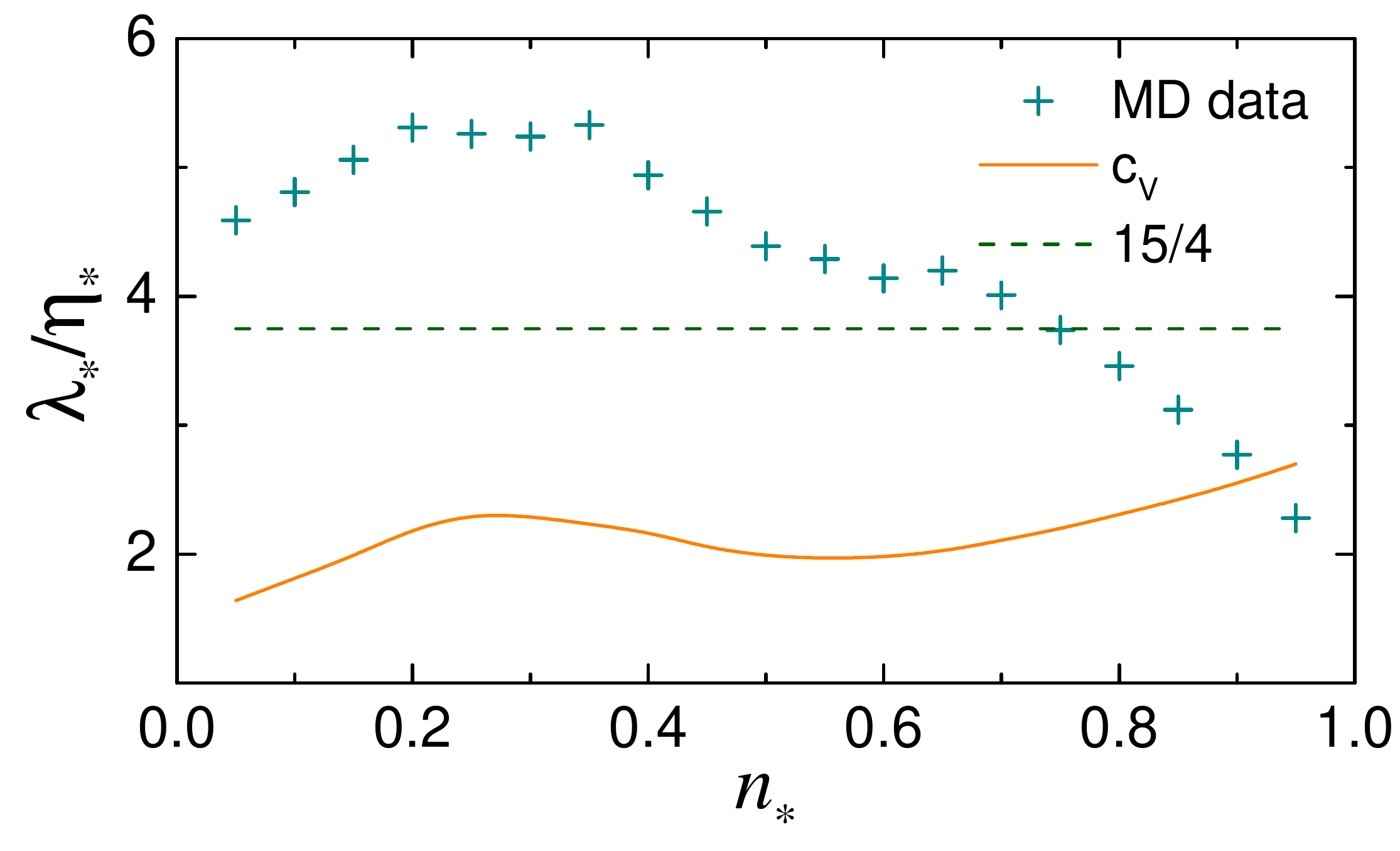}
\caption{The ratio of the reduced thermal conductivity to the viscosity coefficients $\lambda_*/\eta_*$ versus the reduced density $n_*$ in a LJ liquid. Symbols correspond to numerical results tabulated in Ref.~\cite{Meier2002}. The solid curve shows the dependence of $c_V$ on $n_*$. Note a pronounced critical enhancement in the vicinity of the critical density.}
\label{Fig4}
\end{figure}
 
It is tempting to assume that the same vibrational mechanism can be responsible for the momentum transfer in liquids and thus determines their shear viscosity coefficient. Consider a fluid flowing from left to right in the sketch of Fig.~\ref{Fig1} and having a uniform velocity gradient $d u/d x$. By definition, the force between adjacent layers per unit area (the stress) is $\sigma=\eta (d u/d x)$, where $\eta$ is the shear viscosity coefficient~\cite{MarchBook}. On the other hand, the difference in momenta between neighbouring fluid quasi-layers is $m\Delta(d u/d x)$. Vibrating particles transfer this momentum with a characteristic frequency $\langle \omega \rangle/2\pi$. The number of particles per unit area is $\Delta^{-2}$. The force per unit area, related to this vibrational process is $\sigma=m\langle \omega \rangle (d u/d x)/2\pi\Delta$. Combining this with the definition      
of shear viscosity we get
\begin{equation}\label{viscosity}
\eta = \frac{m\langle \omega \rangle}{2\pi\Delta}.
\end{equation}
This essentially coincides with Andrade's point of view on the viscosity of liquids~\citep{Andrade1931,Andrade1934}.

\begin{figure}
\includegraphics[width=8.cm]{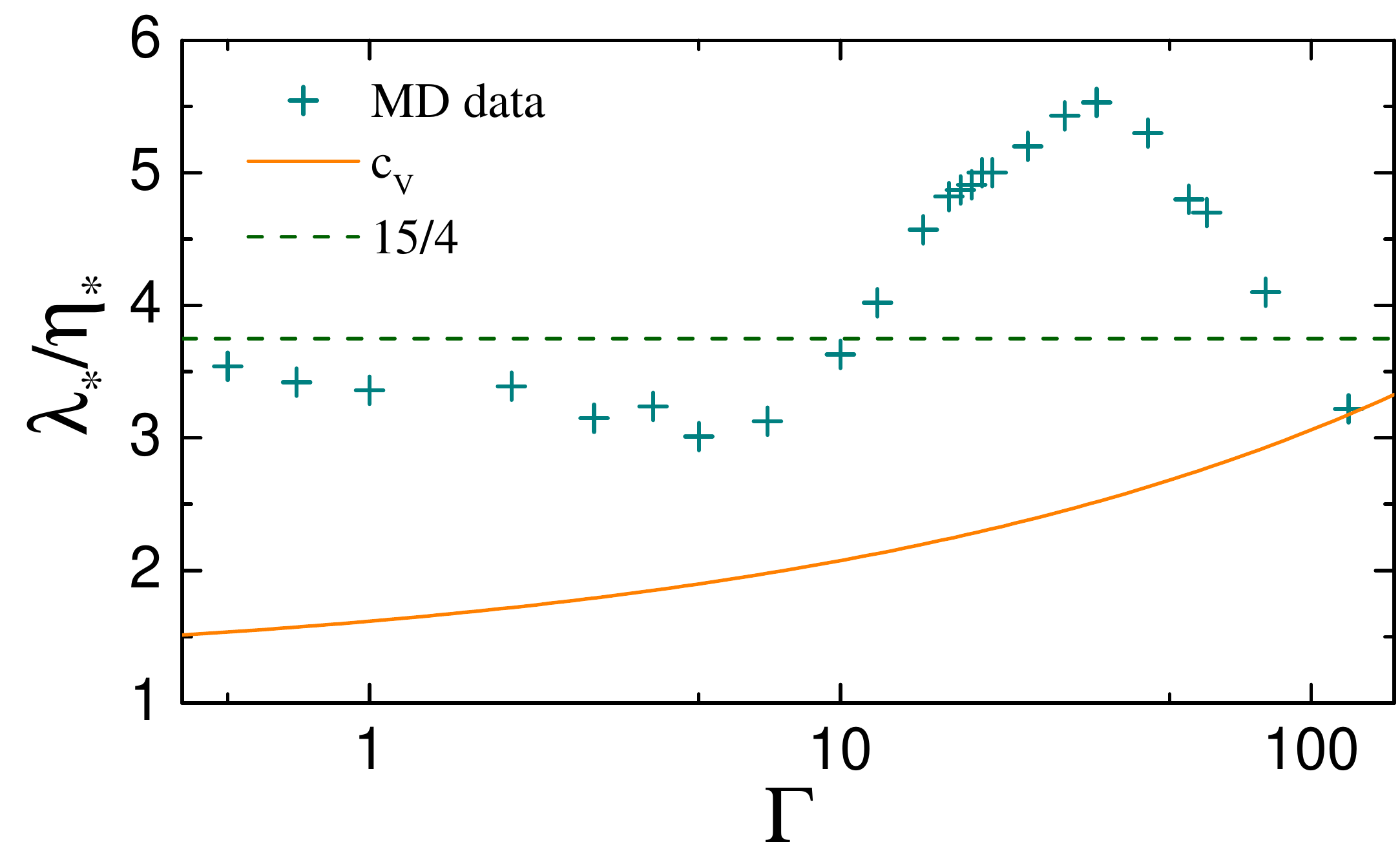}
\caption{The ratio of the reduced thermal conductivity to the viscosity coefficients $\lambda_*/\eta_*$ versus the coupling parameter $\Gamma$ in a OCP fluid. Symbols correspond to numerical results tabulated in Refs.~\cite{ScheinerPRE2019} and~\cite{DaligaultPRE2014}. The solid curve shows the dependence of $c_V$ on $\Gamma$.}
\label{Fig5}
\end{figure}

Comparing equations (\ref{Cond1}) and (\ref{viscosity}) we immediately obtain $\eta=m\lambda/c_V$, or, in appropriately reduced units $\lambda_*/\eta_*=c_V$. In Figure~\ref{Fig4} we plot the ratio $\lambda_*/\eta_*$ in a LJ liquid along the isotherm $T_*=1.35$ as obtained from the numerical simulation~\cite{Meier2002}. The ratio $\lambda_*/\eta_*$ exhibits a pronounced non-monotonous dependence on $n_*$. Although an average value of this ratio predicted by theory is approximately correct ($\simeq c_V$) at high density, the density dependence is not reproduced. Similar picture takes place for the OCP fluid as shown in Fig.~\ref{Fig5}. This provides a strong indication that the mechanisms of momentum and heat transfer in liquids are different. Momentum transfer is not so effective as the vibrational model predicts. This correlates well with the conventional assumption that the mechanisms of mass and momentum transfer in fluids are convective, that is involve atomic hopping (activated jumps) from occupied sites to holes. An average waiting time betwen such rearrangments is considerably longer than the vibrational period. The coupling between the diffusion and viscosity coefficients is also evidenced by the Stokes-Einstein (SE) relation, $D\eta(\Delta/T)\simeq \alpha_{\rm SE}$, where $D$ is the self-diffusion coefficient and $\alpha_{\rm SE}$ is the SE coefficient~\cite{ZwanzigJCP1983}.  The SE relation is satisfied to a very high accuracy in both LJ and OCP with $\alpha_{\rm SE}\simeq 0.15$ for the LJ liquid~\cite{CostigliolaJCP2019,KhrapakMolPhys2019} and $\alpha_{\rm SE}\simeq 0.14$ for the OCP fluid~\cite{DaligaultPRL2006,DaligaultPRE2014}. 

On the other hand, Figs.~\ref{Fig4} and \ref{Fig5} demonstrate that the relation $\lambda_*/\eta_*\simeq 4$ holds with the accuracy of about $50\%$ in the LJ case and about $30\%$ in the OCP case. This ``average'' value is also close to the ideal monoatomic gas limiting result $\lambda_*/\eta_*\simeq 3.75$~\cite{LifshitzKInetics,ONeal1962} (shown by horizontal dashed lines in Figs.~\ref{Fig4} and \ref{Fig5}), which should be appropriate at low densities (weak coupling). Note that the density window corresponding to LJ liquid in Fig.~\ref{Fig4} is about $n_{\rm max}/n_{\rm min}\sim 20$, while for the OCP fluid it is much broader, $n_{\rm max}/n_{\rm min}\sim 10^7$, because $\Gamma\propto n^{1/3}$.     

\section{Conclusion}

To summarize, we have presented a vibrational model of heat transfer in simple liquids with soft interatomic interactions and derived a general expression for the heat transfer coefficient with no free parameters.
The model has been tested on recent accurate MD data on the heat transfer in a dense LJ liquid and a strongly coupled OCP fluid and a remarkably good agreement has been documented. We also demonstrated that a similar mechanism for the momentum transfer in liquids does not lead to satisfactory results for the shear viscosity coefficient, except very near the freezing point.
 
The excellent agreement with MD results for two quite different model systems illustrates the success of the model for soft interaction potentials. The model is likely to become invalid for sufficiently steep hard-sphere-like interactions.
Finding the demarcation line between soft and hard interactions in the context of the vibrational model of heat transfer would be an interesting task for future research (a similar demarcation in the context of instantaneous elastic moduli constitutes an ongoing line of research~\cite{KhrapakJCP2016,KhrapakSciRep2017,KhrapakPRE2019,
KhrapakPRL}.)
It would be also interesting to perform comparison of the theory with numerical and experimental results on other model and real liquids and to look into potential applications to lower space dimensionality. This work is in progress and will be reported elsewhere.


\acknowledgments
I would like to thank Mierk Schwabe for a careful reading of the manuscript.

\bibliography{TC_Ref}

\end{document}